# Anomalous flux quantization and formation of dipole-flux state in a multiply-connected high-$T_c$ NdBa$_2$Cu$_3$O$_{7-\delta}$ superconductor

K. S. Yun[1], I. Iguchi, S. Arisawa, and T. Hatano

*Nano System Functionality Center, National Institute for Materials Science, 1-2-1 Sengen, Tsukuba, Ibaraki 305-0047, Japan*

Abstract

High-resolution scanning superconducting quantum interference device (SQUID) microscopy was used to study the flux quantization phenomenon in multiply-connected anisotropic high-$T_c$ NdBa$_2$Cu$_3$O$_{7-\delta}$ single crystalline thin films. The spatial distribution of internal flux in a hole was found to be non-uniform and changed drastically for applied small fields. With increased fields above 10μT, a local magnetic dipole flux developed inside the hole, in contrast to an isotropic Nb superconductor. The total net flux trapped in a hole was kept to be constant for larger holes, but the abrupt transition of flux quantization state was observed for smaller holes. The possible explanation is given based on the anisotropic d$_{x2-y2}$-wave order parameter of high-$T_c$ superconductors.

One of the most remarkable properties in superconductors is magnetic flux quantization (FQ), i.e. the magnetic flux trapped in a hole in a multiply-connected superconductor with its ring thickness much greater than the penetration depth is quantized in terms of flux quantum unit $\Phi_0 = h/2e = 2.07 \times 10^{-15}$ Wb [1-5]; i.e.

---
[1] Present affiliation: Sony Chemical & Information Device Corporation.

$$\Phi = \oint_C \vec{A} \cdot d\vec{s} = n\Phi_0 \qquad (1)$$

where the loop $C$ is taken deeply inside the superconductor. It depends on neither superconducting material nor the shape of the hole for conventional isotropic superconductors. Theoretically [1-5], it is derived by the condition that the phase change of the order parameter through a certain loop $C$ in a superconductor around a hole should be $2n\pi$ (n: integer), i.e. the wave function should be single valued. The FQ state is rigid and not influenced by a magnetic field below $H_{c1}$. The result only states that the total integrated flux in a hole is $n\Phi_0$ (n=0, 1, 2…) without mentioning any internal flux structure.

The FQ in a hole was discovered about 50 years ago. Two groups performed independently the basic experiment by measuring the inner flux of the superconducting ring and found that only integral numbers of flux quanta were trapped in the hole [6, 7]. For high-$T_c$ superconductors, Gough *et al.* [8] performed measurements using a YBa$_2$Cu$_3$O$_{7-y}$ ceramic ring at the early stage of the high-$T_c$ history and observed small integer numbers of flux quanta jumping in and out of the ring. Later, Kirtley *et al.* observed the quantized flux in a hole and the quantized vortex in superconducting films and crystals [9]. The FQ is considered as an appearance of quantum mechanics in a macroscopic scale.

It is the aim of this paper to study the FQ in a hole in detail including the internal flux structure using a high-resolution scanning SQUID microscopy. The experiment was done by applying a small magnetic field at low temperatures to the sample film with holes of different size in which a few number of magnetic quanta were trapped by the field cooling process. With increasing external magnetic field (much less than $H_{c1}$), the internal magnetic flux became non-uniform in space and a dipole-flux state appeared above a certain field in the hole. The total integrated flux inside the hole was found to be unchanged for the larger holes, but a nearly abrupt change showing a transition of the FQ state was recognized for smaller holes.

The single-crystalline $NdBa_2Cu_3O_{7-\delta}$(NBCO) thin films were prepared on MgO (001) substrates by the Tri-Phase Epitaxy (TPE) method [10,11]. First $BaTiO_3$ and $SrTiO_3$ buffer layers were formed and then the NBCO film of 800nm thick was deposited by TPE process. $T_c$ of the annealed films was 92K. A Nb thin film was prepared by sputtering technique with the thickness of 500 nm on a Si wafer substrate. Square and circular holes with different sizes were patterned in the films by a focused ion beam (FIB) etching technique. DS20 and DS40 are samples with square holes of 20x20μm$^2$ and 40x40μm$^2$ in the NBCO film. DC20 and DC40 are samples with circular holes of ø 20μm and ø 40μm in the NBCO film. SS20 is a square hole of 20x20μm$^2$ in the Nb thin film.

The measurements were carried out by a scanning SQUID microscope (SSM). Many measurements have been performed using SSM [9, 12-18]. The SSM system consists of a dc SQUID, a pickup loop and a scanning system [19]. The dc SQUID made of Nb/AlO$_x$/Nb junctions has the flux sensitivity of 5x10$^{-6}$ $\Phi_0/\sqrt{Hz}$. Both the SQUID and the pickup loop of ø10μm were mounted on a Si chip which was set on a cantilever. The scanning was done with the loop not touching the sample. The sample stage and the sensor stage were shielded by tri-fold permalloy metals. A coil wound around the sample stage generated a local field. The SQUID detects the total magnetic flux penetrating the pickup loop. The scanning direction was along the *a*-axis direction of the NBCO film.

The experiment was performed as follows. First, the sample was cooled down under a low applied magnetic field of a few µT (field cooling process), which led to generate integer multiples of flux quanta $\Phi_0$ inside the holes according as their size. Fig. 1(b)-(c) show the top view of the observed magnetic flux images trapped in the holes with square shape (area: 400 & 900µm$^2$). The vortices present outside the hole are those trapped in the pinning sites during the field cooling process. Each vortex carries one flux quantum $\Phi_0$ and the flat plane (pink-violet color) was identified as the Meissner plane. Spatially flat field distribution was seen inside the hole. The integrated magnetic flux was found to be ~0 and ~1$\Phi_0$ for (b) and (c), respectively.

An external magnetic field along the *z*-direction (*c*-axis) was then increased with the magnetic flux trapped in the hole at 3K. Fig. 2 (a)-(d) show the magnetic images (top view) pictured at different applied magnetic fields from 10µT up to 50µT for sample DS20, and sample DS40, DC40 and sample SS20. As expected, the trapped vortices in the hole ($\cong 2\Phi_0$) in Nb were not changed by increasing the magnetic field, showing that the Meissner effect is perfect. In contrast to the isotropic superconductor Nb, very interestingly, the magnetic images for the NBCO films (DS20, DS40 and DC40) were remarkably varied with increasing field, without any change in the distribution of quantized vortices formed by the field cooling process. With increasing field, the original flat field distribution in DS20 (not shown) inside the hole changed into a lopsided field distribution at 10µT and a clear magnetic dipole flux pair emerged in the hole at 20µT.

Further increase of magnetic field resulted in a remarkable development of a magnetic dipole flux pair peak. The integrated flux around the positive or negative peak attained several times of $\Phi_0$ at 50µT. This structure appeared near lower right and the upper left corners of square in the case of DS20. Fig. 2(b) shows a similar series of magnetic images for DS40. In this picture, the paired structure is principally observed near lower left and upper right corners inside the square hole. Fig. 2(c) shows the case for DC40. The generation of a clear magnetic dipole flux pair emerged again at 20μT. In this case, however, the pair appeared around lower left and upper right corners inside the square hole. The emergence of the magnetic dipole pair were independent of the shape of the hole and started to develop for fields $H_{ex} = 5 - 10$μT as obtained by the data extrapolation process.

It is emphasized that the vortex distribution outside the hole was never disturbed by the external field up to 50μT. The strong magnetic flux expulsion from the NBCO film due to strong pinning effect was observable at the film edges. This fact is consistent with the theoretical calculation based on the potential barrier from geometrical origin [20]. Some elongated structures of trapped vortices irregularly seen in Fig. 2 arose from the mechanical adjustment of a tip at each measurement and were not intrinsic. When the distance between the sample and the pickup loop was adjusted closer, the sensor gave some influence to the isolated vortices along the scanning direction (*a*-axis direction).

Figure 3 (a) shows top view of magnetic images for a dipole flux pair of observed in DS40. Fig. 3 (b) and (c) show the side views of the magnetic image for DS40 when the polarity of applied magnetic field was reversed (+50μT → -50μT). Upon this operation, the polarity of the local dipole flux pair was also inverted, which demonstrates that the external field and the inner flux in the hole are physically connected, that is, the hole is not isolated magnetically. Note that for the Nb film, any change of the inner flux due to inversion of external magnetic field was not recognized.

It was also found that the direction of the dipole flux in the hole depended on the spatial location of the hole in the film as shown in Fig. 3(b). In this figure, the dipole polarity changed across the line through the center. The similar measurement was also performed using the Nb film with exactly the same geometry (multiple holes at the same positions), but no evidence on the formation of dipole flux structure was found at least up to 50μT.

Fig. 4(a)-(c) depict the total integrated flux inside the hole as a function of applied external magnetic field for (a)SS20, (b)DS40 and DC40, (c) DS20 and DC20. The total flux inside the hole was evaluated by a careful analysis on the magnetic distribution data inside and outside a hole by assuming that the trapped vortices in the film carry the flux quantum $\Phi_0$. For Nb samples, it was nearly constant $\cong 2\Phi_0$, hence the FQ rule was kept. For the NBCO film, it was also kept to be constant ($\cong 3\Phi_0$) for the larger holes (square: 1600μm$^2$, circle: 1260μm$^2$) in spite of a drastic emergence of a dipole-flux state in space. For the smaller holes (square: 400μm$^2$, circle: 314μm$^2$), the trapped flux was $2\Phi_0$ or $3\Phi_0$ at 10μT, but it rapidly dropped to zero for higher magnetic field, indicating the transition of FQ states in the hole.

The observed phenomenon is fundamentally quite attractive and the most probable interpretation is given in the following way although the other models are not excluded. The disturbance of the trapped flux inside the hole below $H_{c1}$ at low temperatures is never expected so far as we only deal the solution of London equation with cylindrical symmetry [21]. For high-$T_c$ oxides, it is widely accepted that they have an anisotropic $d_{x2-y2}$-wave pairing symmetry in which a pairing state gives rise to an anisotropic energy gap [9,22-28]. With this symmetry, the pair potential of the form $\Delta_0 \cos 2\theta$ vanishes at the angles $\theta_n = (2n-1)\pi/4$ (n= 1– 4) and the penetration depth becomes infinite in these directions.

There have been experimental and theoretical reports on the penetration depth on the *d*-wave nature of singly-connected bulk superconductors [29-31], which resulted in a small change of the in-plane penetration depth. The calculation was, however, done by integrating over all angles and the experiment was only performed to yield the angle-averaged value of bulk material. There has not been any measurement on the angle-resolved penetration depth. In case that the superconductor contains holes, the singularity on the nodes may intervene strongly because the $\theta_n$ angle direction is well-defined.

Under applied magnetic field, the field gradient (magnetic pressure) between the outer space and the hole with zero or small inner flux will be created along the $\theta_n$ directions (channels). With increasing external fields, the inner flux of the hole is thus affected. In other words, there are four specific channels for magnetic field entry. Although the four-fold symmetry is basically a concept of the momentum space, the presence of a hole in the film enables us to prescribe such specific direction in the real coordinate space through geometrical boundaries, just as the case of anisotropic tunneling in the $d_{x2-y2}$-wave junction [32, 33]. If the channels are Josephson weak-link like, with increasing field, the $\Phi_0$ transition in the hole will occur periodically. On the other hand, if it is completely in the Meissner state, no external field penetrates below $H_{c1}$. We conjecture that the channels correspond to be in some intermediate states between these two physical situations.

Now we consider the channel of magnetic entry with the length $L_n$ along the $\theta_n$ direction (see Fig. 3b). When the external field with upward direction parallel to the *c*-axis is increased at low temperatures, the screening current which has both the radial and $\theta$ components is induced deeply in the specific $\theta_n$ direction so that the screening current near the hole may create the local magnetic field with downward direction inside the hole. Simultaneously, in order to keep the total flux inside the hole constant, another screening current is induced along the opposite counter channel, which creates the local magnetic field with upward direction inside the hole, hence yielding a dipole flux state. The dominant contribution arises from the shortest channel because the magnetic field gradient is the highest there. Referring to Fig. 3b, for the position (1), $L_3$ ($\theta = \frac{5\pi}{4}$) and $L_4$ ($\theta = \frac{7\pi}{4}$) channels are the shortest ones. Since $L_1$ ($\theta = \frac{\pi}{4}$) > $L_2$ ($\theta = \frac{3\pi}{4}$), however, the screening current for the opposite channels is easier to be induced in $L_1$ channel. Thus we consider that the contribution of ($L_3$, $L_1$) channels will be dominant and the ($L_2$, $L_4$) channels act as a secondary contribution. Similarly, for the position (2), the ($L_2$, $L_4$) channels give dominant contribution. The same is said for the position (3). The observed dipole features are quite consistent with this picture.

The abrupt transition of the FQ state seen in Fig. 4(c) may have happened because the two localized fluxes of qualitatively different nature (one is the originally trapped quantized flux, the other is locally created by screening current) interacted more strongly for smaller holes, similar to vortex-vortex repulsion, so that the trapped flux may have been kicked out from the hole. After the flux escape from the hole, no more entry of flux is expected due to the existence of strong screening currents flowing around the hole, hence the $\Phi = 0$ state was kept.

In summary, we have studied the internal magnetic structure inside the hole in a multiply-connected NBCO superconductor. For the samples in the metastable state with a few flux quanta trapped in the holes, by applying a small external magnetic field, the flux distribution in the hole changed greatly. Above 10μT, a clear magnetic dipole flux pair due to screening currents appeared in the hole. The total integrated flux was, however, kept to be constant for larger holes, but it changed nearly abruptly for smaller holes, indicating the transition of the FQ state for fields much smaller than $H_{c1}$. The observed phenomenon is quite essential and may be common among high-$T_c$ $d_{x2-y2}$-wave superconductors. It may be applied to the development of a novel $d_{x2-y2}$-wave device such as a memory.

The authors are very grateful to Profs. Y. Tanaka, M. Ueda, and Dr. X. Hu for invaluable discussions. Dr. Hinode is also acknowledged for supplying Nb films.

(ARISAWA.Shunichi@nims.go.jp)

References

[1] F. London , *Superfluids* (John Wiley & Sons, Inc. New York, 1950), Vol. **1,** p.152.

[2] N. Byers and C.N. Yang, Phys. Rev. Lett. **7,** 46 (1961).

[3] L. Onsager, Phys. Rev. Lett. **7,** 50 (1961).

[4] V. L. Ginzburg, Soviet Phys. JETP **15,** 207 (1962).

[5] F. Bloch and H. E. Rorschach, Phys. Rev. **128,** 1697 (1962).

[6] B. S. Deaver, Jr. and W. M. Fairbank, Phys. Rev. Lett. **7,** 43 (1961).

[7] R. Doll and M. Naebauer, Phys. Rev. Lett. **7,** 51 (1961).

[8] C.E. Gough, M.S. Colclough, E.M. Forgan, R.G. Jordan, M. Keene, C.M. Muirhead, A.I.M Rae, N. Thomas, J.S. Abell, S. Sutton, Nature **326,** 855 (1987).

[9] C. C. Tsuei and J. R. Kirtley, Rev. Mod. Phys. **72,** 969 (2000).

[10] K. S. Yun, *et al*. Appl. Phys. Lett. **80,** 61 (2002).

[11] K. S. Yun, Y. Matsumoto, S. Arisawa, Y.Takano, A. Ishii, T. Hatano, K. Togano, M. Kawasaki, H. Koinuma, IEEE Trans. Appl. Supercond. **13,** 2813 (2003).

[12] C. C. Tsuei, J. R. Kirtley, C. C. Chi, Lock See Yu-Jahnes, A. Gupta, T. Shaw, J. Z. Sun, and M. B. Ketchen, Phys. Rev. Lett. **73,** 593 (1994).

[13] A. Sugimoto, T. Yamaguchi and I. Iguchi, Appl. Phys. Lett. **77,** 3069(2000).

[14] H. J. H. Smilde, Ariando, D. H. A. Blank, G. J. Gerritsma, H. Hilgenkamp, and H. Rogalla, Phys. Rev. Lett. **88,** 057004 (2002).

[15] H. Hilgenkamp, Ariando, HJH Smilde, DHA Blank, G. Rijnders, H. Rogalla, J.R. Kirtley, C.C. Tsuei. Nature **422,** 50 (2003).

[16] I. Iguchi, T. Takeda, T. Uchiyama, A. Sugimoto and T. Hatano, Phys. Rev. B **73,** 224519-1 (2006).

[17] I. Iguchi, T. Yamaguchi, and A. Sugimoto, Nature **412,** 420 (2001).

[18] I. Iguchi, S. Arisawa, K-S. Yun, T. Hatano T. Uchiyama, I. Tanaka, Appl. Phys. Lett. **91,** 202511-1 (2007).

[19] T. Morooka, S. Nakayama, A. Odawara and K. Chinone, Jpn. J. Appl. Phys. **38,** L119 (1999).

[20] E. Zeldov, *et al*. Phys. Rev. Lett. **73**, 1428 (1994).

[21] D. H. Douglass, Jr., Phys. Rev. **132,** 513 (1963).

[22] D. J. Van Harlingen, Rev. Mod. Phys. **67,** 515 (1995).

[23] T. Lofwander, V. S. Shumeiko and G. Wendin, Supercond. Sci. Technol. **14,** R53 (2001).

[24] J. Annet, N. Goldenfeld and A. J. Leggett, Physical Properties of High Temperature Superconductors, Vol.**5**, D. M. Ginsberg (ed.), (World Scientific, Singapore, 1996).

[25] D. A. Wollman, D. J. Van Harlingen, W. C. Lee, D. M. Ginsberg and A. J. Leggett, Phys. Rev. Lett. **71,** 2134 (1993).

[26] D. A. Brawner and H. R. Ott, Phys. Rev. B **50,** 6530 (1994).

[27] I. Iguchi and Z. Wen, Phys. Rev. B**49,** 12388 (1994).

[28] A. Mathai, Y. Gim, R. C. Black, A. Amar and F. C. Wellstood, Phys. Rev. Lett. **74,** 4523 (1995).

[29] W. N. Hardy, D. A. Bonn, D. C. Morgan, Ruixing Liang and Kuan Zhang, Phys. Rev. Lett. **70**, 3999 (1993).

[30] D. Xu, S. K. Yip and J. A. Sauls, Phys. Rev. B **51**, 16233 (1995).

[31] R. Prozorov and R. W. Giannetta, Supercond. Sci. Technol. **19**, R41 (2006).

[32] Y. Tanaka and S. Kashiwaya, Phys. Rev. B**56,** 892 (1997).

[33] W. Wang, M. Yamazaki, K. Lee and I. Iguchi, Phys. Rev. B**60,** 4272 (1999).

Figure captions

Fig. 1 **a** Schematic of a multiply-connected $d_{x^2-y^2}$-wave superconductor containing a square hole. **b**, **c**: Scanning SQUID microscope images of trapped flux in square holes of 20x20μm$^2$ and 30x30μm$^2$ at low background. The quantized vortices around the hole are the trapped ones during the cooling process. The pink-violet region shows the Meissner plane. Almost spatially uniform images inside the holes were obtained. The hole contains zero flux quantum in **b** and ~1$\Phi_0$ in **c.**

Fig. 2 A series of scanning SQUID microscope images of trapped flux in the hole for NBCO and Nb films in the superconducting state at 3K under applied fields from 10μT up to 50μT. **a - c** correspond to the NBCO film with a 20x20μm$^2$ square hole (DS20), a 40x40μm$^2$ square hole (DS40), and a ø40μm circular hole (DC40), respectively. **d** corresponds to the Nb film with a 20x20μm$^2$ square hole (SS20). Note that the background color of Meissner plane changed with field because color only stands relative strength of magnetic field in each measurement. With increasing magnetic field up to 50μT, the magnetic distribution in the hole of NBCO films drastically changed. For example, regarding **a**, almost flat distribution at 1μT (not shown) changed to a lopsided one at 10μT, and a clear dipole-like pair was seen at 20μT, which developed remarkably for further increase of magnetic field. A similar change was observable for **b** and **c** except that their pair positions in the hole are different. In contrast to **a-c**, no change was observable for **d**.

Fig. 3 (a) Top vie of magnetic images for a dipole flux pair when the direction of magnetic field was reversed for the NBCO film with a 40x40μm$^2$ square hole (DS40). (b) Side view of the dipole flux images as the spatial position of holes in the film. (c)The direction of a dipole flux pair was also reversed accordingly, indicating that the external field directly affected the magnetic flux in the hole.

Fig. 4 Total magnetic flux in the hole as a function of applied magnetic field for (a)Nb film with a 20x20μm$^2$ square hole (SS20), (b)NBCO films with square (40x40μm$^2$: DS40) and circular (ø40μm: DC40) holes, (c) and those with square (20x20μm$^2$: DS20) and circular (ø20μm: DC20) holes at 3K. While the total flux inside the hole was kept to be nearly constant for (a) and (b) , it showed a nearly abrupt change of inner flux for (c),   indicating the transition of the FQ state for an applied magnetic field much smaller than $H_{c1}$.

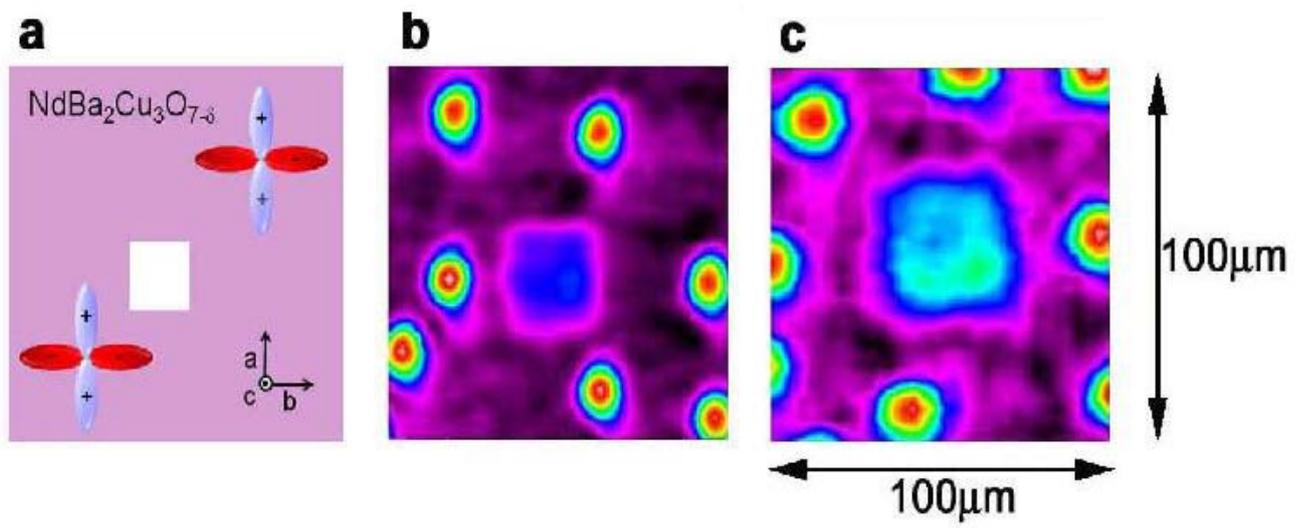

Figure1. K.S. YUN et al.,

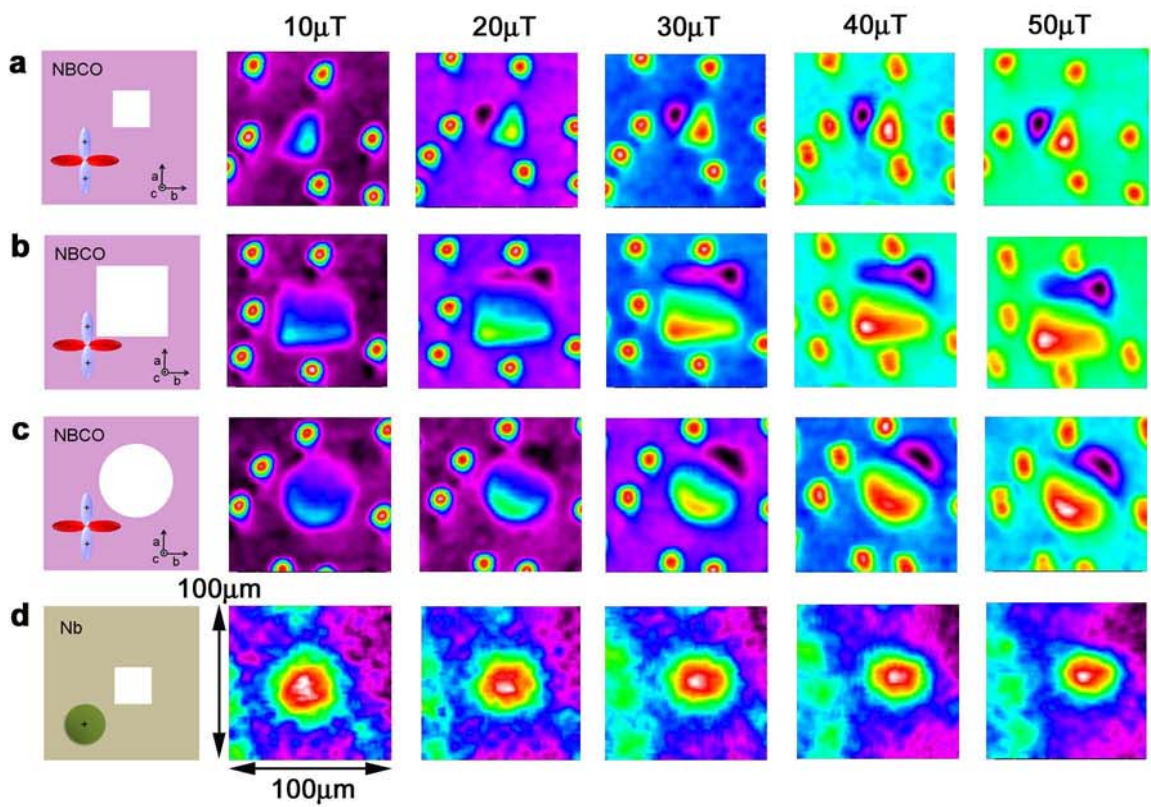

Figure 2. K. S. YUN et al.,

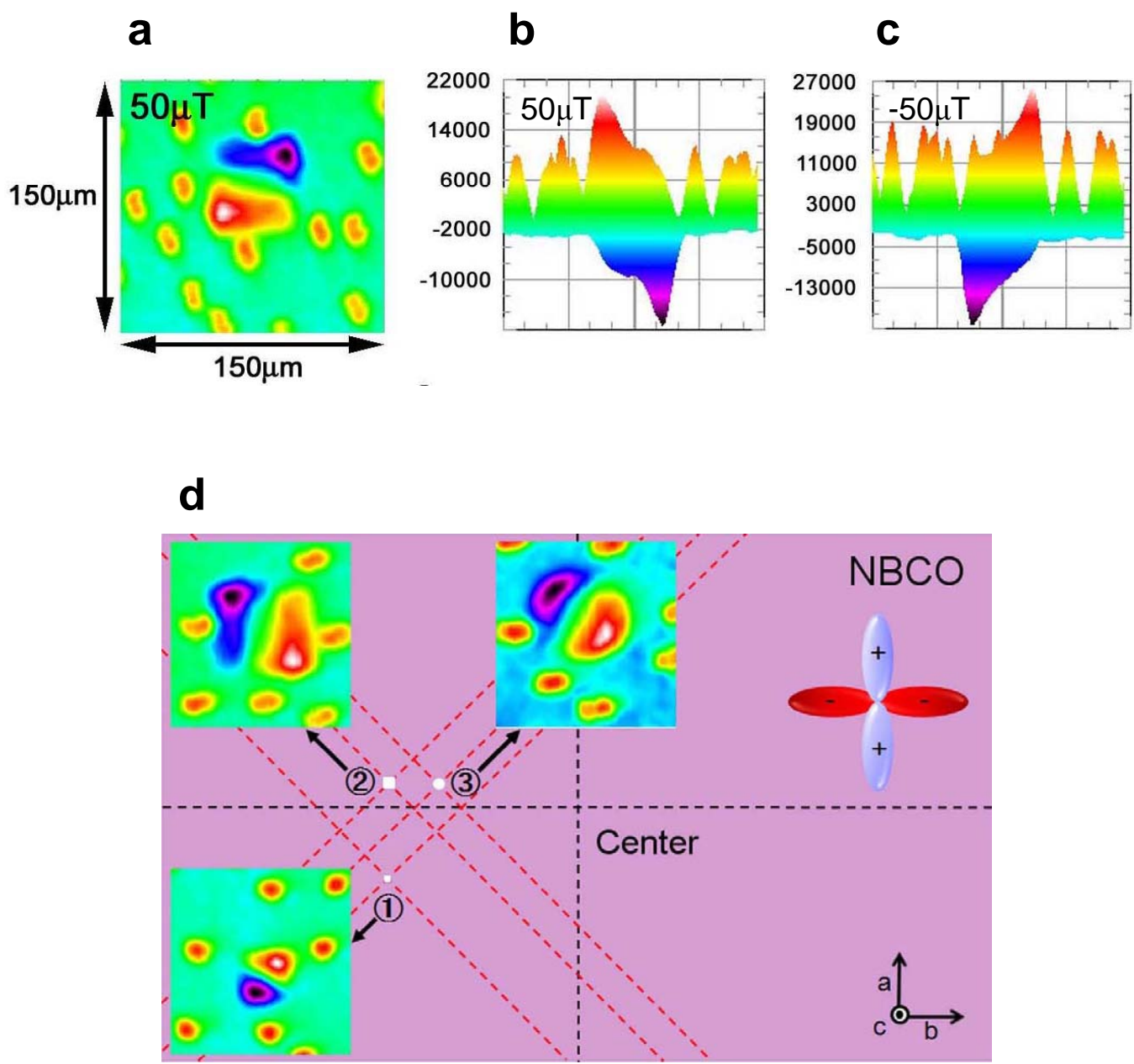

Figure3. K.S. YUN et al.,

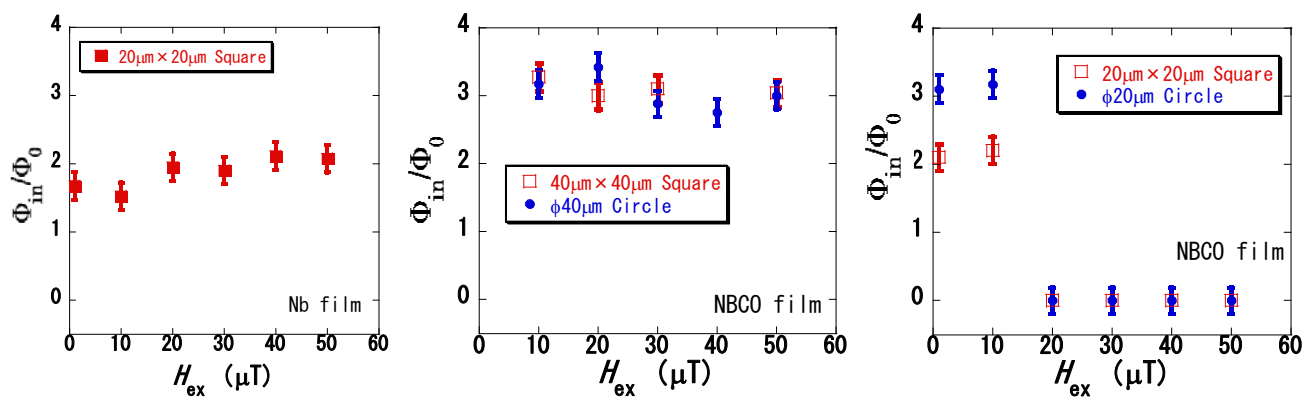

Figure4. K.S. YUN et al.,